\documentclass[12pt,fleqn]{article}
\usepackage{amssymb}
\usepackage{amsmath}

\newcommand{\be} {\begin{equation}}
\newcommand{\ee} {\end{equation}}

\newcommand{\ba} {\begin{array}}
\newcommand{\ea} {\end{array}}
\newcommand{\bd}{\begin{displaymath}}
\newcommand{\ed}{\end{displaymath}}

\newtheorem{theorem}{Theorem}

\begin{document}
\large
\begin{center}
\large {\bf On the Symmetry Approach to Reduction  \\ of
Partial Differential Equations  }
\end{center}
\begin{center}
{\large  I.Tsyfra  \footnote{ E-mail: tsyfra@math.uwb.edu.pl}}
 \end{center}
\begin{center}
{\it  University of Bialystok, 2 Academicka str., Bialystok, Poland;\\
 Intitute of Geophysics, 32 Palladina av., 03142, Kyiv,
Ukraine;}\\
 
\end{center}

\begin{abstract}
 \noindent 
 We  propose the symmetry reduction method of partial differential
equations to the system of differential equations with fewer number of
independent variables. We also obtain genera\-lized sufficient conditions for
the solution found by conditional symmetry method to be an invariant one in
classical sense.
\end{abstract}

\begin{center}
1.INTRODUCTION
 \end{center}

In recent years the symmetry method is often used for reduction of partial
differential equations to the equations with fewer number of independent
variables and thus for construction of exact solutions for different
mathematical physics problems. To construct a corresponding ansatz
generators of classical Lie point transformations are used as well as
operators of conditional symmetry. In this connection the application
of combination of conditional and generalized symmetry is fruitful as was
shown in \cite{Tsyfra:fokas&liu,Tsyfra:zhdanov}
on the examples of evolution equation in two-dimensional case
(see also \cite{Tsyfra:tsyfra99}). In \cite{Tsyfra:svirshchevskii}
   Svirshchevskii proposed the symmetry reduction method
based on the invariance of linear ordinary differential equations
(see also \cite{Tsyfra:kamran&milson&olver}). It is the symmetry explanation
of ``nonlinear'' separation of variables \cite{Tsyfra:galaktionov}
for the evolution type equations.

Here we propose an approach applicable for symmetry reduction of partial
differential equations which are not restricted to evolution type ones.
It can be used in multi-dimensional case. This approach is the
generalization of the method introduced in \cite{Tsyfra:svirshchevskii}.

\section{Generalized symmetry and reduction\\ of partial differential
equations }

Let us consider partial differential equation
\begin{equation}
U ( x, u,{\mathop u\limits_1},{\mathop u\limits_2},\ldots ,
{\mathop u\limits_k} ) =0,  \label{tsyfra:1}
\end{equation}
where $x=(x_1,x_2,\ldots ,x_n)$, $u=u(x)\in C^k(\mathbb{R}^n,{\mathbb
R}^1)$, and
${\mathop u\limits_k}$ denotes all partial derivatives of $k$-th order.
Replacing $u$ by $u+\epsilon w$ and then equating coefficients
 at $\epsilon$ in Taylor series expansion we obtain linearized
 equation~(\ref{tsyfra:1})
\begin{equation}
L(x, u, w)=0. \label{tsyfra:2}
\end{equation}
It has been proved that the following property is fulfilled in this case.
If equation~(\ref{tsyfra:1})
admits Lie--B\"acklund vector field $Q=\eta ( x, u,{\mathop u\limits_1},
{\mathop u\limits_2},\ldots ,{\mathop u\limits_r} ) \partial_u$ and
$u=f(x)$ is a solution of equation~(\ref{tsyfra:1}) then
\begin{equation}
w=Qu|_{u=f(x)}     \label{tsyfra:3}
\end{equation}
is a solution of equation~(\ref{tsyfra:2}).

This property is illustrated by the connection between the solutions of
Liouville and Moutard equations. It is well known that the Liouville equation
\begin{equation}
u_{xy}=2\exp u     \label{tsyfra:4}
\end{equation}
is invariant with respect to the Lie group of transformations with generator
\[
Q_1=f(x)\partial_x+g(y)\partial_y-(f'+g')\partial_u,
\]
where $f(x)$ and $g(y)$ are arbitrary smooth functions. Then from the solution
\[
u= \ln \frac{X' Y'}{(X+Y)^2}
\]
of Liouville equation we easily obtain the solution
\[
u=\frac{X'_1}{X'}+\frac{Y'_1}{Y'}-2\frac{X_1+Y_1}{X+Y}
\]
of the Moutard equation
\[
w_{xy}=2\frac{X'Y'}{(X+Y)^2} w
\]
with potential $V=2\frac{X'Y'}{(X+Y)^2}$, where $X(x)$, $Y(y)$ are
arbitrary smooth functions of their arguments, $fX'=X_1$, $gY'=Y_1$,
by using (\ref{tsyfra:3}).

This property can also be used for reduction of partial differential equations
to system of equations with smaller number of independent variables.
For simplicity consider ordinary differential equation
\begin{equation}
H (x, u,\ldots ,
\underset{m'}{u} )=0,  \label{tsyfra:5}
\end{equation}
where $\underset{m'}{u}$ denotes the derivative of $u$ with respect
to one variable $x_1$ of $m'$-th order, and $H$
are a smooth function of its arguments. Suppose that $Q$ is the operator
of Lie--B\"acklund symmetry  of equation~(\ref{tsyfra:5}). Let
\begin{equation}
u=F(x, C_1,\ldots,C_{m'}),     \label{tsyfra:6}
\end{equation}
where $F$ is a smooth function of variables $x, C_1,\ldots ,C_{m'}$,
$C_1,\ldots, C_{m'}$
are arbitrary functions of parametric variables $x_2, x_3,\ldots, x_n$,
be a general solution of equation~(\ref{tsyfra:5}).
 Then operator $Q$ transforms solution (\ref{tsyfra:6}) to the solution of
 linearized version of equation~(\ref{tsyfra:5}) i.e.  the linear
 homogeneous ordinary differential equation.  It means that $Q$ maps
 the set of solutions (\ref{tsyfra:6}) into a $m'$--dimensional  vector space
 $M$. Moreover, we proved that partial derivatives
 $\frac{\partial F}{\partial C_i}$, $i=\overline{1,m'}$ form the basis of
  $M$ provided that $H$ and $F$ are sufficiently smooth functions
  of their arguments. In this connection the following statement holds.
\begin{theorem}\label{tsyfra:theorem1}
Let equation~(\ref{tsyfra:5}) be invariant with respect to
the Lie--B\"acklund operator $Q$. Then the  ansatz
\begin{equation}
u=F(x, \phi_1, \phi_2,\ldots,\phi_{m'}),     \label{tsyfra:7}
\end{equation}
where $\phi_1, \phi_2,\ldots,\phi_{m'}$ depend on $n-1$ variables
 $x_2, x_3,\ldots, x_n$ reduces partial differential equation
\begin{equation}\label{tsyfra:8}
\eta (x, u,\underset{1}{u},
\underset{2}{u},\ldots,\underset{r}{u} )=0
\end{equation}
to the system of $k_1$ equations for unknown functions
 $\phi_1, \phi_2,\ldots ,\phi_m$ with $n-1$ independent variables
 and $k_1 \le m'$.
\end{theorem}

Taking into account the above--mentioned arguments we proved the theorem.
It can be easily generalize for $\phi_1, \phi_2,\ldots ,\phi_{m'}$ depending
on $\omega_l(x)$, $l=\overline{1, n-1}$, where $\omega_l(x)$ are some 
functions of variables $x$. Note that
the case when equation~(\ref{tsyfra:5}) is linear ordinary differential
equation and $\eta$ is the function of special type (evolution type)
 was considered in \cite{Tsyfra:svirshchevskii}.

We consider several examples illustrating the application of the theorem.
Firstly consider equation
\begin{equation}
u_t=f(u_x)u_{xx},     \label{tsyfra:9}
\end{equation}
where $f(u_x)$ is a smooth function of $u_x$. Let us consider the
equation~(\ref{tsyfra:5}) in the following form
\begin{equation}
u_{xx}=u_x^3.     \label{tsyfra:10}
\end{equation}
In accordance with our approach we study the invariance of the
equation~(\ref{tsyfra:10}) with respect to Lie--B\"acklund operator
\begin{equation}
K_1=(u_t-f(u_x)u_{xx})\partial_u.     \label{tsyfra:11}
\end{equation}
Function $f(u_x)$ is determined from the condition of invariance
of the equation~(\ref{tsyfra:10}) with respect to the
operator~(\ref{tsyfra:11}).It has the form
\[
f=\frac{A}{u_x^3}+\frac{B}{u_x^2},
\]
where $A$, $B$ are arbitrary real constants.
Then we proved that the equation~(\ref{tsyfra:10}) admits the operators
\[
K_2=u u_x\partial_u, \qquad K_3=h(u+u_x^{-1})\partial_u,
\]
where $h$ is arbitrary smooth function.
Therefore the equation~(\ref{tsyfra:10}) is invariant with respect to
the group of
Lie--B\"acklund transformations with infinitesimal operator $\alpha_1 K_1+
\alpha_2 K_2+\alpha_3 K_3$, where $\alpha_1$, $\alpha_2$, $\alpha_3$
are arbitrary real constants.
From this it follows that the theorem can be used for the equation
\begin{equation}
u_t=  (\frac{A}{u_x^3}+\frac{B}{u_x^2} )u_{xx}+
\lambda uu_x+ \lambda_1 h(u+u_x^{-1}),  \label{tsyfra:12}
\end{equation}
where $\lambda$, $\lambda_1$ are arbitrary real constants.
The ansatz corresponding to the equation~(\ref{tsyfra:10}) has the form
\begin{equation}
u=\phi_2(t)-({\phi_1(t)-2x})^{1/2},     \label{tsyfra:13}
\end{equation}
where $\phi_1(t)$ and $\phi_2(t)$ are unknown functions. The
ansatz~(\ref{tsyfra:13}) reduces the equation~(\ref{tsyfra:12}) to the
system of ordinary differential equations
\begin{gather}
\phi_2'=A-\lambda +\lambda_1 h(\phi_2),\qquad
-\phi_1'=2(B+\lambda \phi_2).  \label{tsyfra:14}
\end{gather}
For different $h(\phi_2)$ we receive different solutions. Let $h(\phi_2)=
\phi_2$. In this case the solution of the system~(\ref{tsyfra:14}) is
\[
\phi_2 = Ce^{\lambda_1t} - (A-\lambda)\lambda^{-1}_1 ,\quad
-\phi_1 = 2(B-(A-\lambda)\lambda \lambda^{-1}_1)t+
2\lambda^{-1}_1\lambda Ce^{\lambda_1t}-C_1.  \label{tsyfra:15}
\]
Substituting this solution into (\ref{tsyfra:13}) we obtain the solution
\[
u=Ce^{\lambda_1t} - (A-\lambda)\lambda^{-1}_1 -
[2((A-\lambda)\lambda \lambda^{-1}_1-B)t
-2\lambda^{-1}_1\lambda Ce^{\lambda_1t}+C_1 -2x]^{1/2}
\]
of equation~(\ref{tsyfra:12}). This solution can not be obtain by using
classical Lie method of point symmetry.

Note that the ansatz~(\ref{tsyfra:13}) can  be used for reduction
of equations obtained by commutating of operators  $K_1$, $K_2$,
$K_3$ also.

Next we show the application of the method to equation associated
with inverse scattering problem. As we know the group--theoretical
background of the inverse scattering problem method
was given for the first time in \cite{Tsyfra:fushchych&nikitin}.

Namely we study the symmetry of the linear
ordinary differential equation
\begin{equation}
u_{xx}=f(t,x)u.   \label{tsyfra:16}
\end{equation}
where variable $t$ play the role of parameter in this equation,
with respect to the Lie--B\"acklund operator
\begin{equation}
Q_1= (u_t+u_{xxx}-3\frac{u_{xx}u_x}{u}+\alpha (t)u )\partial_u,   \label{tsyfra:17}
\end{equation}
where $\alpha(t)$ is a function of $t$.

We have proved that the equation~(\ref{tsyfra:16}) admits operator
(\ref{tsyfra:17}) if and only if $f$ satisfies the Korteveg-de Vrise
equation in the form
\begin{equation}
f_t+f_{xxx}-6ff_x=0.   \label{tsyfra:18}
\end{equation}
This statement is valid for arbitrary smooth $\alpha(t)$. Thus if one can
construct the general solution of equation~(\ref{tsyfra:16}) for some
solution $f(t,x)$ of (\ref{tsyfra:18}) then the corresponding ansatz will
reduce partial differential equation
\begin{equation}
u_t+u_{xxx}-3\frac{u_{xx}u_x}{u}+\alpha (t)u=0  \label{tsyfra:19}
\end{equation}
 to the system of two ordinary
differential equations with independent variable $t$.
We stress that solution $f(t,x)$ should not necessarily vanish at the infinity.

To solve the Cauchy problem
\begin{equation}
u\mid_{t=t_0}=g(x)  \label{tsyfra:20}
\end{equation}
for equation~(\ref{tsyfra:19}) one should construct solution $f=p(t,x)$ of
equation~(\ref{tsyfra:18}) satisfying the condition
$p(t_0,x)=\frac{g_{xx}}{g}$ and then to integrate ordinary differential
equation
\[
u_{xx}=p(t,x)u.
\]
If we can construct the ansatz in this way then we reduce the Cauchy
problem (\ref{tsyfra:20}) for the equation~(\ref{tsyfra:19}) to the
Cauchy problem for system of two ordinary differential equations.

In addition note that theorem 1 helped us to generalize theorem proved in
\cite{Tsyfra:tsyfra2002}
and concerning the sufficient conditions for the solution obtained by using
conditional symmetry operators to be an invariant solution in the classical
sense. Namely consider involutive family of operators
\begin{equation}
Q_a=\xi_{aj}(x,u)\partial_{x_j}+\eta_a(x,u)\partial_u, \qquad
a=\overline{1,m}.               \label{tsyfra:21}
\end{equation}
We have summation on repeated indeces.Suppose that the equation~(\ref{tsyfra:1}) is conditionally invariant with
respect to involutive family of operators~(\ref{tsyfra:21}) and
corresponding ansatz reduces
this equation to ordinary differential equation of $k_1$-th order.
Then the following statement holds.
\begin{theorem}\label{tsyfra:theorem2}
Let equation~(\ref{tsyfra:1}) be invariant with respect to
$s$--dimensional Lie algebra $AG_s$ and conditionally invariant with
respect to
involutive family of operators $\{Q_i\}$. If the system
\[
\xi^l_i \frac{\partial u}{\partial x_l}=\eta_i(x,u)
\]
is also invariant under the algebra $AG_s$ and $s\ge k_1+1$,
then conditionally invariant solution of equation~(\ref{tsyfra:1})
with respect to involutive family of operators $\{Q_i\}$ is an invariant
solution in the classical Lie sense.
\end{theorem}
It is necessary to note that this theorem can be generalize to the case when
the algebra $AG_m$ contains Lie--B\"acklund operators too.

\section{Conclusion}

We showed that the symmetry of ordinary differential equations
can be used for reduction of partial differential equations to the system
with fewer number of independent variables.
 It follows from Theorem~1 that the linearity is not the necessary
 condition for this reduction.
It is obvious that the suggested method is applicable in many-dimensional
case.

In addition note that this approach can be used in solving the
problem of integrability of partial differential equations. It gives
the possibility to reduce this problem to the problem of integrability
for ordinary differential equation. We consider this method to be important
in studding quasi-exactly solvable systems.

We also obtained the generalized sufficient condition for the solution
constructed by using conditional symmetry to be an invariant one in classical
sense.


\end{document}